\documentclass[preprint,showpacs,preprintnumbers,amsmath,amssymb]{revtex4}

\usepackage{graphicx}% Include figure files
\usepackage{dcolumn}% Align table columns on decimal point
\usepackage{bm}% bold math

\def\nn{\nonumber}
\def\sg(#1){\textrm{sign}(#1)}

 \def\ap{{\alpha'}}

\def\cP{\mathcal{P}}
\def\cI{\mathcal{I}}
\def\cN{\mathcal{N}}
\def\I{\underline{\textrm{I}}}
\def\J{\underline{\textrm{J}}}
\def\m{\textrm{mass}}
\def\REF#1{(\ref{#1})}
\begin{document}

\preprint{\vspace{-1cm}IPhT-T08/57}
%\preprint{\vspace{-1cm}arXiv:YYMM.NNNvV}
\title{\ \\
Absence of Triangles in Maximal Supergravity Amplitudes\\
\ \\ }% Force line breaks with \\

\author{N.~E.~J.~Bjerrum-Bohr}
\email{bjbohr@ias.edu}
\affiliation{%
Institute for Advanced Study,\\
School of Natural Sciences, Einstein Drive, Princeton, New Jersey
08540, USA
}%
\author{Pierre Vanhove}
\email{pierre.vanhove@cea.fr}
\affiliation{
Institut de Physique Th\'eorique,\\
CEA, IPhT, F-91191 Gif-sur-Yvette, France\\
CNRS, URA 2306, F-91191 Gif-sur-Yvette, France\\ \ \\ }%

\date{\today}% It is always \today, today,

\begin{abstract}
From general arguments, we show that one-loop $n$-point amplitudes
in colourless theories satisfy a new type of reduction formula.
These lead to the existence of cancellations beyond supersymmetry.
Using such reduction relations we prove the no-triangle hypothesis
in maximal supergravity by showing that in four dimensions the
$n$-point graviton amplitude contain only scalar box integral
functions. We also discuss the reduction formulas in the context of
gravity amplitudes with less and no supersymmetry.
\end{abstract}

\pacs{04.60.-m, 04.65+e}% PACS, the Physics and Astronomy
\keywords{Supergravity}%Use showkeys class option if keyword
\maketitle

\section{\label{sec:level1}Introduction}
The knowledge of perturbative gravity amplitudes and their
UV-behaviour is to a large extent based on arguments from power
counting rather than on explicit computations. Since the calculation
of graviton scattering amplitudes is a lengthy and challenging
subject -- accurate power counting arguments incorporating all
symmetries of the amplitude, {\it e.g.} gauge symmetry,
supersymmetry etc, can be a preferred option. Much care should
however be taken not to overlook symmetries that could drastically
reduce an expected power counting behaviour.

The one-loop $n$-graviton amplitude in a Feynman diagram approach in
$D=4-2\epsilon$ dimensions takes the generic form
\begin{eqnarray}
\label{e:Mn1} M_{n ;1} \, &=& \mu^{2\epsilon}\, \int {d^D \ell\over (2\pi)^D}\,
{\prod_j^{2n} (q_{\mu_j}^{(2n,j)}\ell^{\mu_j}) + \prod_j^{2n-1}
(q_{\mu_j}^{(2n-1,j)}\ell^{\mu_j})
+ \cdots + K \over \ell_1^2\, \cdots \ell_n^2} \\
\nonumber & \equiv& \mu^{2\epsilon}\, \int {d^D \ell\over (2\pi)^D}\, {\cP_{n}
(\ell)\over \ell_1^2\, \cdots \ell_n^2}\,.
\end{eqnarray}
Here $\ell_i^2 =  (\ell - k_1 - \cdots -  k_i)^2$ are the
propagators along the  loop and $q_{\mu_j}^{(i,j)}$ are functions of
external momenta  and polarisations.  The integration is  performed
by separating~\cite{Bern:1993kr} the  four dimensional dependence
and the  $D-4=-2\epsilon$   part  of  the  loop   momentum,
$\ell=\bar \ell+\ell_{[-2\epsilon]}$ with
$\ell^2_{[-2\epsilon]}=\mu^2$ given by the infrared regulator so
that
\begin{equation}
  \int  {d^D\ell\over  (2\pi)^D}=  \int {d^4\bar\ell\over  (2\pi)^4}\,
  \int {d^{-\epsilon}(\mu^2)\over (2\pi)^{-2\epsilon}}\,.
\end{equation}

Momentum conservation implies that $k_1+\cdots +k_n=0$. $K$ is a
constant. Because  of the two derivative nature of the cubic
gravitational   coupling, the numerator $\cP_{n} (\ell)$ is a
polynomial with at most $2n$ powers of loop momentum $\ell$.
Supersymmetry can be accounted for by a correcting factor of
$\ell^{-\cN}$, in a theory with $\cN$ on-shell supercharges. The
maximal order for the numerator polynomial for the $n$ point
amplitude in eq.~(\ref{e:Mn1}) is thus $\ell^{2n-\cN}$ in this power
count.

According to the above na\"\i ve powercounting the $n$-graviton
one-loop amplitude in $\cN=8$ supergravity has at most $2(n-4)$
powers of $\ell$ in the numerator polynomial.
The loop momentum polynomial can be manipulated via a succession of
(ordered) integral reductions cancelling one power of loop momentum
at each step of reduction~\cite{intred,Campbell:1996zw,
Bern:1993kr,Denner:2005nn}. This reduces all tensor integrals in the
amplitude expression into a linear combination of scalar integrals.
The na\"\i ve power counting in ${\cal N}=8$ supergravity indicates
that one-loop amplitudes should be expandable in a basis of {\it
scalar box, triangle and bubble integral functions} as well as {\it
rational pieces}.

Surprisingly in a number of explicit calculations using on-shell
unitarity
techniques~\cite{Dunbar,Bern:1998sv,Bern:2005bb,BjerrumBohr:2005xx,BjerrumBohr:2006yw,
Bern:2007xj}  it   has  however been  observed that this na\"\i ve
power count does not reflect reality and  that  maximal supergravity
amplitudes have a much better power  counting.  This has also been
referred   to as the ``no-triangle''     hypothesis of ${\cal N} =
8$ supergravity~\cite{BjerrumBohr:2006yw,Bern:2007xj}. The
``no-triangle'' hypothesis suggests that $\cN=8$  supergravity
amplitudes in  four dimensions are completely specified by a basis
of box integral functions and in particular do neither contain {\it
triangles} or {\it bubble functions}  nor {\it rational pieces}. By
analysing the  dependence on the  loop momenta in the cuts it was
deduced in~\cite{Bern:2007xj} that $n-4$ powers of loop momenta must
have cancelled in  the total amplitude in comparison to the counting
in eq.~(\ref{e:Mn1}),
 yielding the ``no-triangle'' hypothesis.

 In this paper we demonstrate that for colourless gauge theories, like
 gravity,  there are  new  reduction formulas  in  place for  on-shell
 amplitudes  where at  each step of reduction \emph{two  powers  of loop
 momentum are cancelled}. This contrasts integral reduction formulas
 for ordered theories where only {\it one power of loop
 momentum is cancelled}. The unordered integral reductions are instrumental
 in exhibiting  the improved divergence structure of colourless gauge
 theories  such as  gravity. We  will  first discuss  on very  general
 grounds  how  a set  of  reduction  formulas  for unordered  integral
 functions at  one-loop can be  induced by gauge invariance.   Next we
 will discuss  the consequences  for maximal supergravity  and explain
 why  the na\"\i  ve power  counting  is  incorrect  and how  the  generic
 $n$-graviton  amplitude in  four dimensions  in  maximal supergravity
 cannot  contain any basis integral other than  scalar
 box integral functions. This argument follows directly from an application
 of the new integral reduction formulas on the generic graviton
 amplitude where polarisation tensors has been expanded in terms of
 the momenta of the external legs.
 This constitute a direct proof of the ``no-triangle'' hypothesis for
 ${\cal N} = 8$ supergravity.

%%%%%%%%%%%%%%%%%%%%%%%%%%%%%%%%%%%%%%%%%%%%%%%%%%%%%%%%%%%%%%%%%%%%%%%%%%%%
\section{\label{sec:level1b}One-loop  gravity amplitude in  the string
based formalism}
%%%%%%%%%%%%%%%%%%%%%%%%%%%%%%%%%%%%%%%%%%%%%%%%%%%%%%%%%%%%%%%%%%%%%%%%%%%%
In this section we will review the string based formalism to be
employed later in the paper for proving the ``no triangle''
hypothesis of the $\cN=8$ $n$-graviton amplitudes at one-loop
following~\cite{StringBased,Strassler:1992zr,Dunbar,BjerrumBohr:2008vc}.

The string based formalism is very natural in theories with no
colour factors since it in a simple way incorporates expressions for
one-loop amplitudes which has complete crossing symmetry of all
external legs.  The crossing symmetry implies that the various
colour ordered gravity amplitudes have the same tensorial structure
and imply additional simplicity.

We will first consider an $n$-point one-loop $\varphi^3$ scalar
field theory amplitude. We will introduce Feynman parameters $a_i$
and exponentiate the propagators. Through this we can write the
ordered scalar $n$-point one-loop $\varphi^3$ amplitude as
\begin{eqnarray}
\nn I_n&=&\mu^{2\epsilon} \int {d^D\ell\over \pi^{D\over2}}
\prod_{i=1}^n {1\over (\ell-k_{1}-\cdots
    -k_{i})^2}\\
\nn  &=&   \mu^{2\epsilon}\,  \int  {d^D\ell\over   \pi^{D\over2}}
\, \prod_{i=1}^n  \int_0^\infty d\alpha_i \,
\exp\left(-\sum_{i=1}^n \alpha_i
(\ell-k_1- \cdots -k_i)^2\right)\\
  &=&  \mu^{2\epsilon}\,  \int  {d^D\hat\ell\over \pi^{D\over2}}\,
e^{-T\, \hat \ell^2}\, \int_0^\infty {dT\over T}\, T^{-n}\,
\prod_{i=1}^n   \int_0^1   da_i  \,\delta\left(1-\sum_{i=1}^n
a_i\right)\, \exp(-T\, Q_n)\,.
\end{eqnarray}
Here we have used $\mu$ to denote the infrared regulator. We have
exponentiated the propagators by introducing the Schwinger parameter
$\alpha_i$. The Feynman parameters have next been rescaled by the
proper-time of the loop defined as $T=\alpha_1+\cdots +\alpha_n$
according to $a_i=\alpha_i/T\in [0,1]$ for $1\leq i\leq n$. We have
used the following definitions~\cite{Bern:1993kr}
\begin{equation}\label{defQnA}
Q_n=\sum_{1\leq i<j\leq n}\, S_{ij} \, a_ia_j\,,
\end{equation}
where
\begin{equation}
S_{ij}\equiv{1\over 2}\, (k_i+\cdots +k_{j-1})^2, \qquad i\neq j, \
S_{ii}\equiv 0\,.
\end{equation}
Performing the integration over the shifted loop momentum
\begin{equation}
\label{e:ellhat} \hat \ell = \ell - K_{[n]}\,, \qquad K_{[n]}=
\sum_{i=1}^n k_i \,\sum_{j=1}^i a_i\,,
\end{equation}
we obtain the following expression,
\begin{equation}
\label{e:orderd}
I_n=     \Gamma\left(D-1\over2\right)\,     \int_0^\infty     {dT\over
  T}\,T^{-D/2+n}\, \int_0^1 da_n\int_0^{a_n}da_{n_1}\cdots
\int_0^{a_2}da_{1} \,\delta(1-\sum_{i=1}^n a_i)\, e^{-T\, Q_n}\,.
\end{equation}
We will remark that if one performs the integration over the
proper-time $T$ then the familiar expression for an ordered
$n$-point one-loop scalar amplitude as used in~\cite{Bern:1993kr} is
obtained
\begin{equation}
I_n =\Gamma\left(D-1\over 2\right) \Gamma\left(n-{D\over2}\right)
\int_0^1 da_n\int_0^{a_n}da_{n_1}\cdots
\int_0^{a_2}da_{1}\,\delta(1-\sum_{i=1}^n a_i)\,
Q_n^{{D\over2}-n}\,.
\end{equation}
If we perform a change of variables according to
\begin{equation}\label{e:nutoa}
\nu_i= \sum_{j=1}^i \, a_j\,,
\end{equation}
we can rewrite the expression for $Q_n$ in eq.~(\ref{defQnA}) in the
following way
\begin{equation}\label{defQn}
Q_n=\sum_{1\leq i<j\leq n}\, (k_i\cdot k_j)\,
\Big[(\nu_i-\nu_j)^2-|\nu_i-\nu_j|\Big]\,.
\end{equation}
This will be the expression that we will use in the main part of the
paper. In the ordered amplitude eq.~(\ref{e:orderd}) all the $\nu_i$
are ordered according to $0\leq \nu_1\leq \nu_2\leq \cdots \leq
\nu_n$. Using the $\nu_i$ variables the total momentum along the
loop is given by the expression
\begin{equation}\label{e:KnDef}
K_{[n]}= \sum_{i=1}^n k_i \, \nu_i\,.
\end{equation}
In this representation the $\nu_i$  are  the  dual variables to the
loop momenta and any power of $\nu_i$ in the integrand of the
integral represents powers of the loop momentum in the amplitude.

The unordered amplitudes can be obtained by summing the expression
eq.~(\ref{e:orderd}) over all orderings of the external legs
\begin{eqnarray}
\nn
\mathcal{I}_n&=&\mu^{2\epsilon} \sum_{\sigma\in\mathfrak{S}_n}\int
{d^D\ell\over \pi^{D\over2}} \prod_{i=1}^n {1\over (\ell-k_{\sigma(1)}-\cdots
-k_{\sigma(i)})^2}\\
\label{e:scalar}&=&\Gamma\left(D-1\over 2\right) \int_0^\infty
{dT\over T}\, T^{-D/2+n}\, \int_0^1 d\nu_1\cdots d\nu_{n}\, {1\over
n} \left[\sum_{i=1}^n \delta(\nu_i=1)\right]\, e^{-T \, Q_n}\,.
\end{eqnarray}
Here  $\mathfrak{S}_n$ is  the set  of  all permutations  of the $n$
orderings of the external legs.  Because of the  summation over all
orderings of the external legs, the $\nu_i$ variables with  $1\leq
i\leq n$ are freely integrated between $[0,1]$. By translational
invariance along the loop and momentum conservation the integration
depends  only on $n-1$ variables. We can fix this translational
invariance around the loop by freezing one of the $\nu_i$ to 1. In
order to preserve the symmetry among the $\nu_i$ variables we insert
a symmetrised delta-function.

For an ordering of the external  legs specified  by  a permutation
$\sigma\in\mathfrak{S}_n$ of $n$  objects, the  mapping between the
$\nu_i$ and  $a_i$ variables takes the form
\begin{equation}
\nu_i= \sum_{j=1}^i \, a_{\sigma(j)}\,.
\end{equation}
The  absolute  value in  $Q_n$ in eq.~(\ref{defQn}) allows us to
formally break up the domain of integration for the integral into
various regions of analyticity in the complex energy momentum plane.
These regions can be seen to correspond exactly to the possible
physical orderings of the external legs. For instance, for the case
of the four-point amplitude the integral $\cI_4$ becomes
\begin{eqnarray}
\nn  \cI_4&=&\Gamma\left(D-1\over2\right)\,     \int_0^\infty    {dT\over
    T}\, T^{-D/2+4}\, \int_0^1d\nu_1\cdots d\nu_3\, \delta(\nu_4=1)\,
  e^{-T\,Q_4}\\
&=& I_4(s,t)+I_4(s,u)+I_4(t,u)\,,
\end{eqnarray}
where $I_4(s,t)$ is the scalar  box function evaluated in the physical
region $s,t<0$
\begin{equation}
  I_4(s,t)={1\over st}\, \left[{2\over \epsilon^2}\,
  \left((-s)^{-\epsilon}+(-t)^{-\epsilon}\right)
  -\ln^2(-s/-t) -\pi^2+ \mathcal{O}(\epsilon)\right]\,,
\end{equation}
with identical definitions for $I_4(s,u)$ and $I_4(t,u)$.

For interacting colourless gauge theories like gravity the general
structure of the one-loop $n$-point amplitude in $D$ dimensions is
given by~\cite{StringBased,Strassler:1992zr,Dunbar}
\begin{equation}\label{e:AmpDef}
\mathcal{M}_{n;1}= \int_0^\infty {dT\over T}\, T^{-D/2+n}\, \int_0^1
d\nu_1\cdots  d\nu_{n}\, {1\over n} \left[\sum_{i=1}^n
\delta(\nu_i=1)\right]\, \cP(h_{r_is_i},k_i;\nu_i)\,e^{-T\,Q_n}\,.
\end{equation}
Here $h_{r_is_i}$ is the polarisation tensor of the $i$th graviton
with momentum $k_i$. Comparing with eq.~(\ref{e:scalar}) the new
ingredient is the quantity $\cP(h_{r_is_i},k_i;\nu_i)$ arising from
the three, four and higher point interaction vertex that encodes the
tensorial structure of the amplitude.

One important property  of the representation~(\ref{e:AmpDef}) is
that the  orderings  of  the  external legs  all have  the
\emph{same} tensorial structure. This  is a consequence of  the form
of  the expression for $\mathcal{M}_{n;1}$  where  the polarisation
dependence  enters  the function   $\cP(h_{r_is_i})$ which   is
integrated   over   the unconstrained variables $[0,1]$.

As  in  the  case  of  the  scalar amplitude that we described
earlier the various regions of  analyticities in the  complex energy
plane arise from  expanding  the absolute  value constraints in
$Q_n$ and the sign function  $G_F$.

For $n$-graviton one-loop amplitudes in supergravity theories with
$\cN$ on-shell supersymmetries (counting the number of
supersymmetries  in units of  four dimensional Majorana
supercharges) with $0\leq \cN\leq 8$, one way to derive the
coefficient $\cP(h_{r_is_i})$ is to extract it from the following
generating function
\begin{equation}\label{e:P}
      \cP(h_{r_is_i},k_i;\nu_i)=\prod_{I=1}^{\cN/4}\prod_{m=1}^4 d\psi^m_I d\bar\psi^m_I\prod_{\alpha,\dot\alpha=1}^{2}
      \prod_{i=1}^n         \int         d\theta^{\alpha }_i
      d\bar\theta^{\dot\alpha}_i e^{\sum_{i=1}^n      (\theta^1_i      \,
h_i\cdot\psi+\theta_i^2 \,(ik_i\cdot \psi))+c.c.} e^{{\cal
        F}+\bar{\cal F}}\Big|_{\textrm{multilinear}}\,.
 \end{equation}
In this equation $\theta_i^{\alpha}$ and $\dot\theta_i^{\dot\alpha}$
are $\alpha,\dot\alpha=1,2$ anticommuting variables and  $\psi_I^m$
and $\bar\psi_I^m$  are $1\leq m\leq  4$ and $1\leq I,J\leq \cN/4$
space-time fermionic zero modes (counting the number of
supersymmetries in units of four dimensional Majorana supercharges).

In this expression one has to extract the  multilinear part  in the
polarisations  $h_i$ and $\bar h_i$ with $1\leq i\leq n$ where we
have decomposed the polarisations of the graviton $h_{r_is_i}$ as a
symmetric product of two spin one  polarisations $h_{r_i}$ and $\bar
h_{r_i}$, {\it i.e.}, $h_{r_is_i}=(h_{r_i}\bar h_{r_i}+h_{r_i}\bar
h_{s_i})/2$. The expression  for $\exp({\cal     F})$ is given
by~\cite{StringBased,Strassler:1992zr,Dunbar,BjerrumBohr:2008vc}
\begin{eqnarray}
\nn
{\cal F}&=&
{T\over2}\sum_{i\neq j} (h_{i}\cdot h_{j}) \,\theta_i\theta_j\, \partial_{i}\partial_{j}G_B(\nu_i-\nu_j)
        +{i\over2}\sum_{i\neq    j}     (k_{i}\cdot
h_{j}\,\theta_i-k_{j}\cdot h_{i}÷\,\theta_j) \partial_{i}G_B(\nu_i-\nu_j)\\
\nn &+& {1\over2}\sum_{i\neq j} (h_{i}\cdot h_{j}) \,
G_{F}(\nu_{i}-\nu_j)
 -{i\over2}\sum_{i\neq j} (k_{i}\cdot h_{j}\, \theta_{j}-k_{j}\cdot h_{i}\, \theta_{i})\, G_{F}(\nu_i-\nu_j)\\
&+&{1\over2}\sum_{i\neq j} \theta_{i}\theta_{j}\,(k_{i}\cdot
k_{j})\,      G_{F}(\nu_i-\nu_j)\,. \label{e:Feven}\end{eqnarray}
One can use an  equivalent  definition  for  $\bar{\cal F}$  where
all the polarisations are taken to be  $\bar h_i$, and the fermionic
variables $\bar\theta_i$. We have introduced a scalar world-line
Green's function  $G_B(x)$ and the $G_F(x)$ function defined by
\begin{equation}\label{e:Prop}
  G_B(x) = x^2 -|x|,\qquad G_F(x)=\sg(x)\,,
\end{equation}
so that
\begin{equation}\label{e:GBQ}
  \sum_{1\leq i<j\leq n} \, (k_i\cdot k_j)\, G_B(\nu_i-\nu_j)=
  Q_n\,.
\end{equation}
The origin of  each term in this expression can be  traced back to
the various contractions between the word-sheet variables. [The
following definitions can be obtained by considering the $\ap\to0$
limit of  the corresponding superstring quantities as  detailed in
the appendix  of~\cite{BjerrumBohr:2008vc}. They  can be  derived as
well without any reference to string theory and are obtainable by
considering the Green's function in the worldline
formalism~\cite{StringBased,Strassler:1992zr}].

\medskip \noindent{$\bullet$\ \bf The bosonic contractions for $i\neq
j$}\vspace{-0.1cm}\smallskip
\begin{eqnarray}
\nn  \langle h_i\cdot \partial X(\nu_i) \, h_j\cdot
 \partial      X(\nu_j)\rangle       &=&      2(h_i\cdot      h_j)\,
 (\delta(\nu_i-\nu_j)-1)\,,\\
\label{e:HigherD1} \langle h_i\cdot \partial X(\nu_i)\,\bar h_j\cdot
\bar\partial X(\nu_j)\rangle&=&  2(h_i\cdot \bar h_j)\,
(\delta(\nu_i-\nu_j)-1)\,,\\
\nn\langle h_i\cdot \partial X(\nu_i)\,k_j\cdot
 X(\nu_j)\rangle &=&(h_i\cdot k_j)\, \partial_{\nu_i}G_B(\nu_i-\nu_j)\,.
\end{eqnarray}
\smallskip\medskip\noindent{$\bullet$} {\bf The fermionic contractions for
$i\neq j$}\vspace{-0.1cm}\smallskip
\begin{equation}
\label{eq:hhF1} \langle v_i\cdot \psi(\nu_i) \, v_j\cdot
\psi(\nu_j)\rangle= (v_i\cdot v_j)\, G_F(\nu_i-\nu_j)\,.
\end{equation}
Here $v_i$ stands for an external polarisation $h_i$ or momentum
$k_i$. We will return to the structure  of one-loop amplitudes in
supergravity theories in section~\ref{sec:level4}.

When extracting  the multilinear part  in the polarisations  $h_i$
and $\bar h_i$ which define $\cP(h_i,\bar h_j,k_i,\nu_i)$ in
eq.~(\ref{e:P}) one finds that the result is expressible solely in
terms of the first derivative  of   the   Green   function
$\partial_xG_B(x)$ and the second derivative of the Green function
\begin{eqnarray}\label{e:zm}
\partial^2_x G_B(x)=2\,(\delta(x)-1)\,.
\end{eqnarray}

We want to stress that to the contrary
of~\cite{StringBased,Strassler:1992zr,Dunbar} we are \emph{not}
systematically integrating out  the double  derivatives of the Green
function $G_B$.

We will make use of an expansion of the polarisations of the
external states in a basis of independent momenta
\begin{equation}\label{e:cij1}
h_i=\sum_{j=1}^{n-1} c_i{}^j \, k_j+ q^\perp\,.\vspace{-0.2cm}
\end{equation}
Here $q^\perp$ is a vector orthogonal to the $(n-1)$ linearly
independent  external momenta. The momentum $k_j$ is only needed for
dimensions $D>4$. An identical definition for the $\bar h_i$
polarisations can be employed. Using the relation eq.~(\ref{e:GBQ})
one easily derives that
\begin{eqnarray}
 \sum_{j=1}^n (h_i\cdot k_j)\, \partial_i G_B(\nu_i-\nu_j)&=&
\sum_{j=1}^n\, (h_i\cdot k_j)\,[-2  \, \nu_j - G_F(\nu_i-\nu_j)]\\
 \nn&=&\sum_{r=1}^{n-1}  c_i{}^r\, \left[\partial_r  Q_n+  \sum_{j=1}^n (k_r\cdot
k_j) \, (G_F(\nu_r-\nu_j)-G_F(\nu_i-\nu_j))\right]\,,
\end{eqnarray}
which implies that  $\cP$ is only a function of the first
derivatives of $Q_n$ and $G_F$. Hence the amplitude takes the
symbolic form
\begin{equation}
  \label{e:cPsusy}
\mathcal{M}_{n;1}^\cN=  \sum_{r+s+u=2n-\cN\atop 0\leq u\leq
n}\sum_{l=0}^u\, t^l_{r,s}\,\cI^{[D+2(u-l)]}_{n-l}[\I_r,\J_s]\,.
\end{equation}

It is clear that  the integration over the fermionic variables in
eq.~(\ref{e:P}) will bring powers of the polarisations multiplied by
the $G_F$ function in eq.~(\ref{e:Prop}) or the second derivative of
the propagator $\partial_x^2G_B$ in~(\ref{e:zm}). Thereby the number
of powers of single derivatives on the propagators are reduced. This
reduces the number of single derivatives of $Q_n$ via the relation
in eq.~(\ref{e:GBQ}). The integration over the $4\cN$ fermionic
variables in eq.~(\ref{e:P}) implies the following restriction on
the powers $\partial Q_n$ and $G_F$ in eq.~(\ref{e:cPsusy})
\begin{equation}\label{e:rsConstraint}
r+s\leq 2n -\cN\,.
\end{equation}
The  cancellation of  $\cN$  powers  of  loop momenta  in the $n$
point one-loop amplitude  depends on the  number of fermionic zero
modes which are independent of the nature of the external states. As
a result of this an amplitude with other external states than
gravitons from the massless supergravity multiplet will take the
same form as in eq.~\eqref{e:cPsusy}, since   the  integrals
$\cI_n[\I_r,\J_s]$ that  arise   from  the correlators of the
world-line fields in eq.~\eqref{e:HigherD1} and eq.~\eqref{eq:hhF1}
have the same form independently of the external massless states.

Because of  the   relation  between   the  loop   momentum  and the
total momentum $K_{[n]}$ defined
in eq.~(\ref{e:KnDef}) as well as the relation
\begin{equation}\label{e:QtoK}
  \partial_{\nu_i}Q_n=-2\,k_i\cdot K_{[n]}-  \sum_{m=1}^n \, (k_i\cdot
  k_m)\, G_F(\nu_i-\nu_m)\,,
\end{equation}
the constraint in eq.~(\ref{e:rsConstraint})  is equivalent to  the
statement that one-loop $n$-graviton  amplitudes with $\cN$
supersymmetries have the maximum power of loop momentum given by
\begin{equation}\label{e:Plim}
\cP_{n}(h_i,\bar h_i,k_i;\nu_i) \ \sim \ \ell^{2n-\cN}, \quad
\textrm{for}\quad \ell\gg 1\,.
\end{equation}
This cancellation  arises from the  saturation of the $\cN$  zero
modes for the fermions running in the loop and is independent of the
number of external  states and the dimension as long as the number
of supersymmetries are preserved. The constraint in
eq.~(\ref{e:rsConstraint}) will turn to be important for the
analysis leading to the proof of the ``no-triangle'' hypothesis of
the $\cN=8$ supergravity amplitude in section~\ref{sec:level4}.

We    have   not    integrated   over    the    one-loop proper-time
in eq.~(\ref{e:AmpDef}) because  the contributions from first  term
of the first  line  in eq.~(\ref{e:Feven})  will  give rise to
expressions with higher powers of $T$ to be integrated over. This
will give rise two different   types   of    contributions. The
delta-function part from eq.~(\ref{e:zm}) will give rise to an
amplitude with less points and the constant piece in~(\ref{e:zm})
will give rise to dimension shifted integrals. These will be
discussed in detail in section~\ref{sec:dimshift}.

%%%%%%%%%%%%%%%%%%%%%%%%%%%%%%%%%%%%%%%%%%%%%%%%%%%%%%%%%%%%%%%%%%%%%%%%%%%%
\section{\label{sec:level2}Reduction formulas}
%%%%%%%%%%%%%%%%%%%%%%%%%%%%%%%%%%%%%%%%%%%%%%%%%%%%%%%%%%%%%%%%%%%%%%%%%%%%

In this section we will derive new integral reduction identities
relevant for the colourless $n$-point one-loop amplitude. We will
define the
 \emph{unordered} $n$-point integral in $D$ dimensions
\begin{equation}\vspace{-.1cm}
\label{e:DefIn} {\cal I}_n[\I_r,\J_s] \equiv  \int_0^1\!d^{n-1}\nu
\, Q_n^{D/2-n} \prod_{i\in \I_r} \partial_{\nu_i}Q_n
\prod_{x\in\J_s} G_F(x)\,. \vspace{-.1cm}\end{equation}
These integrals will appear generically in amplitudes in the context
of a string based
formalism~\cite{StringBased,Strassler:1992zr,BjerrumBohr:2008vc}. In
this paper the amplitude  are evaluated in $D=4-2\epsilon$
dimensions.

In the above integral formula the  proper-time has been integrated
out and all orderings of the external legs are integrated over. We
have defined the set of indices $\I_r=\{i_1,\dots,i_r\}$ with $i_1$,
\dots,  $i_r$  indices taking values in $\{1,\dots,n\}$. We have
defined the set of the differences of positions
$\J_s=\{\nu_{j_1}-\nu_{k_1},\dots,\nu_{j_s}-\nu_{k_s}\}$ with $j_1$,
\dots, $j_s$ and $k_1$, \dots, $k_s$ indices taking values in
$\{1,\dots,n\}$ which are the arguments of the sign function
$G_F(x)$ in the integral.
The integration is performed over the position $\nu_i$ of the
external legs along the loop with the measure\vspace{-0.2cm}
\begin{equation}
d^{n-1}\nu= d\nu_1\cdots d\nu_n \, {1\over n}\, \sum_{i=1}^n
\delta(\nu_i-1)\,.\vspace{-0.1cm} \end{equation} The translational
invariance around the loop is fixed by freezing one of the $\nu_i$
to 1. In order  to preserve the symmetry along the $\nu_i$ variables
we insert a symmetrised delta-function. [For a
given ordering, say the canonical ordering $(123\cdots n)$, the
$\nu_i$ are related to the (rescaled) Feynman parameters $a_i$ used
in ref.~\cite{Bern:1993kr} by the linear relations
$\nu_i=\sum_{1\leq m \leq i} a_m$ given in~(\ref{e:nutoa}).]

\noindent The quantity $Q_n$ is defined by\vspace{-0.1cm}
\begin{equation}
\label{eq:Qn} Q_n\equiv \sum_{1\leq i<j \leq n} \, (k_i\cdot k_j)\,
G_B(\nu_i-\nu_j)\,. \vspace{-.15cm}\end{equation}
The functions $G_B(x)$ and $G_F(x)$ are defined by\vspace{-0.1cm}
\begin{equation}
G_B(x)= x^2-|x|\,, \ \ \ \ \  G_F(x)=\sg(x)\ .
\vspace{-0.1cm}\end{equation}
[These quantities are the infinite tension limit ($\alpha'\to0$)
bosonic and fermionic string propagators of the genus one amplitude
(supersymmetry does however not play a r\^ole
here)~\cite{StringBased,Strassler:1992zr,BjerrumBohr:2008vc}.]

\medskip
\noindent{$\bullet$} We will begin by considering $n$-point
integrals with $r+1$ insertions of $\partial Q_n$ factors and no
insertions of $G_F$ functions, so that
$\cI_n[\I_{r+1}]\equiv\cI_n[\I_{r+1},\emptyset]$\vspace{-.1cm}
\begin{equation}\label{e:DefInnoJs}
\cI_n[\I_{r+1}]=\! \int_0^1 \!\!\! d^{n-1}\nu\, Q_n^{D/2-n} \!\!\!
\prod_{i\in \I_{r+1}}\!\!\! \partial_{\nu_i}Q_n\,,
\vspace{-.2cm}\end{equation}
where $\I_{r+1}\equiv\{i_1,\dots,i_{r+1}\}$. Now assume that
$i_{r+1}$ has multiplicity $m$ in $\I_{r+1}$, {\it i.e.}
$\I_{r+1}=\{i_{r+1}\}^{m}\cup \I_{r-m+1}$ with
$i_{r+1}\not\in\I_{r-m+1}$. We will set as well
$\I_r=\I_{r-m+1}\cup\{i_{r+1}\}^{m-1}$,
$\I_{r-1}^{(r+1)}=\I_{r-m+1}\cup\{i_{r+1}\}^{m-2}$, and finally
$\I^{(j)}_{r-1}\equiv \{i_1,\dots,\hat\imath_j,\dots,i_{r-m+1}\}\cup
\{i_{r+1}\}^{m-2}$, {\it i.e.} the list of $r-1$ indices with $i_j$
omitted. Using that $Q_n^{D/2-n}\partial_{\nu} Q_n = (D/2-n+1)^{-1}
\partial_{\nu} Q_n^{D/2-n+1}$
and integrating by parts in eq.~(\ref{e:DefInnoJs}) we get
that\vspace{-0.1cm}
\begin{eqnarray}
 \cI_n[\I_{r+1}]&=&\nonumber\\
\label{e:R1}&& \hspace{-2.3cm}{2\over
D/2\!-\!n\!+\!1}\!\Bigg[\!\!\!\!\!\!\!\!\!\!\!\!\!\!\!\sum_{\ \ \ \
\ \ \ \  \ j\in \I_{r-m+1}} \!\!\!\!\!\!\!\!\!\!\!\!\!\!
(k_{i_{r+1}}\cdot k_j)
\! \left(\!-\cI_{n-1}^{(i_{r+1}j)}[\I^{(j)}_{r-1}]+\cI_n^{[D+2]}[\I^{(j)}_{r-1}]\right)\\
\nonumber &&\hspace{0.2cm}+(m-1)\sum_{s=1}^n (k_{i_{r+1}}\cdot
k_s)\,\cI_{n-1}^{(i_{r+1}s)}[\I^{(r+1)}_{r-1}]\Bigg]\,.
\label{e:Step1}\vspace{-.4cm}\end{eqnarray}
Because  of the  relation eq.~(\ref{e:QtoK}) a power  of
$\partial_{\nu}Q_n$ corresponds    to   a power    of    loop
momenta $\ell$    since $\partial_{\nu_i}Q_n\sim k_i\cdot\ell$. The
above relation expresses the $n$-point amplitude $\cI_n[\I_{r+1}]$
with $r+1$ insertions of loop momentum factors
  as  a  linear   combination  of   the  $n-1$-point
amplitudes     with    $r-1$     insertions     of    loop     momenta
$\cI_{n-1}^{(i_{r+1}j)}[\I^{(j)}_{r-1}]$                            and
  $\cI_{n-1}^{(i_{r+1}s)}[\I_{r-1}^{(r+1)}]$. We have
\begin{equation}
\!  \cI_{n-1}^{(ij)}[\I^{(j)}_{r-1}]\!\equiv\! \int_0^1d^{n-1}\nu\,
  Q_n^{D/2-n}\, \delta(\nu_i-\nu_j)\! \prod_{s\in
  \I^{(j)}_{r-1}}\!\!
  \partial_{\nu_s}Q_n\,,
\vspace{-0.2cm}\end{equation} and the $n$-point dimension shifted
  integral $\cI_n^{[D+2]}[\I^{(j)}_{r-1}]$  evaluated in dimensions
  $D+2$ is
 \begin{equation}
   \cI^{[D+2]}_n[\I_r]=\int_0^1     d^{n-1}\nu     \,    Q^{(D+2)/2-n}
   \prod_{i\in \I_r} \partial_{\nu_i}Q_n\,.
 \end{equation}
 Integrals with more than one mass are
defined  in the same  way with several delta function insertions.
The integration by part produces a  boundary term\vspace{-0.1cm}
\begin{equation}\begin{split}
\partial \cI_n[\I_{r+1}]&= {1\over D/2-n+1}\,\int_0^1d^{n-1}\nu\,
Q_n^{D/2-n+1}\cr &\hspace{-1.3cm}\times
\left[\delta(\nu_{i_{r+1}}=1)-\delta(\nu_{i_{r+1}}=0)\right]
      \, \prod_{i\in \I_r} \partial_{\nu_i}Q_n\,,
\vspace{-.2cm}\end{split}\end{equation}
which is vanishing because of the 1-periodicity of $Q_n$ in each of the
$\nu_i$ variables, $Q_n(\nu_1,\dots,\nu_i+1,\dots)=Q_n(\nu_1,
\dots,\nu_i,\dots)$ since $G_B(1-x)=G_B(x)$ for
$0\leq x\leq 1$ and $G_B(0)=G_B(1)=0$.

\smallskip

These reduction formulas hence express any $n$-point integrals with
$r$ powers of loop momenta  \emph{summed over all orderings of the
external legs} as linear combinations of $(n-1)$-point one-mass
integrals with $r-2$ powers of loop momenta and possibly dimension
shifted integrals. This can be summarised by the schematic
rule\vspace{-.1cm}
\begin{equation}
\label{e:Red1} \cI_n[(\partial Q_n)^r]\rightsquigarrow
\cI^\m_{n-1}[(\partial Q_n)^{r-2}]+ \cI_n^{[D+2]}[(\partial Q_n)^{r-2}]\,.
\vspace{-.2cm}\end{equation}

\noindent{$\bullet$} When some factors of $G_F(x)$ are present in
the integrand we have to distinguish between the following
cases\vspace{-0.3cm}
\begin{itemize}
\item[$\triangleright$] If all the $i\in \I_r$ are such that $\nu_i$ is not an
argument of $G_F(x)$  for any  $x\in \J_s$,  then the  same
manipulations leading to eq.~(\ref{e:Step1}) and eq.~(\ref{e:Red1})
apply with no changes.
\item[$\triangleright$] If   $i_{r+1}$ has multiplicity one in
$\I_{r+1}=\I_r\cup \{i_{r+1}\}$ with $i_{r+1}\not\in \I_r$ and
$\J_1=\{\nu_{i_{r+1}}-\nu_j\}$ then
\end{itemize}\vspace{-0.5cm}
\begin{eqnarray}\label{eqnuGF}
&&\cI_n[\I_{r+1},\J_1]=
{1\over
  D/2-n+1}\!\int_0^1\!\!d^{n-1}\nu\,\partial_{\nu_{i_{r+1}}}Q_n^{D/2-n+1}
%\\ &&\qquad\qquad\qquad\qquad\times
\, G_F(\nu_{i_{r+1}}-\nu_j)\,
\prod_{i\in\I_r}
\partial_{\nu_i}Q_n\,.
\end{eqnarray}
Integrating this  by parts leads to\vspace{-.2cm}
\begin{eqnarray}
\cI_n[\I_{r+1},\J_1]&=&{
2\over  D/2-n+1}\times\\
\nonumber   &&\hspace{-2.5cm}\Bigg[   \sum_{j\in\I_r}
(k_{i_{r+1}}\cdot
  k_j) \nonumber
(-\cI_{n-1}^{(i_{r+1}j)}[\I_{r-1}^{(j)},\J_1]+\cI_n^{[D+2]}[\I_{r-1}^{(j)},\J_1])\\
\nonumber&&\hspace{-0.5cm}+ \Big((n-1)\cI_{n-1}^{(i_{r+1}
j)}[\I_{r}]-\sum_{l=1}^n \cI_n^{(i_{k+1}l)}[\I_{r}^{(r+1)}]\Big)\Bigg]\,.
\vspace{-.1cm}\end{eqnarray}

This is easily generalised to other cases, with higher multiplicity
of $i_{k+1}$ and with additional $G_F$ contributions.
We have the special cases,
\begin{eqnarray}
\nonumber \cI_n[\{i\},\{\nu_r-\nu_s\}]&=&0\,,\ {\rm for} \ i\not\in\{r,s\}\,,\\
\label{e:noHD} \cI_n[\{i\},\{\nu_i-\nu_j\}]&=&        -{2\over
D/2-n+1}\,
\cI^{(ij)}_{n-1}[\emptyset]\,, \\
\nonumber\cI_n[\emptyset,\{\nu_r-\nu_s\}]&=&0\,.
\end{eqnarray}
An   obvious   generalisation   of   these   identities   imply that
$\cI_n[\partial  Q_n,   (G_F)^s]=0$ for $s\geq2$ and
$\cI_n[(G_F)^s]=0$ for all $s>0$.

We can conclude that when some $G_F$ factors are present at best
only \emph{one power of loop momentum} is cancelled at each step of
reduction as in the usual integral reductions
case~\cite{intred,Campbell:1996zw,Bern:1993kr, Denner:2005nn}. This
can be summarised schematically by the rule
\begin{equation}
\label{e:Red2}\begin{split} \cI_n[(\partial Q_n)^r,G_F]&\ \rightsquigarrow\
\cI_{n-1}^\m[(\partial Q_n)^{r-1}]+ \cI_{n-1}^\m[(\partial Q_n)^{r-2},G_F]%\cr
%&\hspace{2.7cm}
+\cI_{n}^{[D+2]}[(\partial Q_n)^{r-2},G_F]\,.
\end{split}\end{equation}
We have shown that in the unordered integral, because of the sum
over all the permutations of the external legs, new integral
reduction formulas, given schematically by eq.~(\ref{e:Red1}) and
eq.~(\ref{e:Red2}), are valid.

Before closing this section we will make a few remarks concerning
the reduction formulas.

The main  reason for  introducing the integrals  in
eq.~(\ref{e:DefIn}) is because they  are the  building blocks of the
`string  based' method. The  ordered integrals considered
in~\cite{Bern:1993kr} are expressed in terms of  the Feynman
parameters $a_i$ which  are linearly related  to the $\nu_i$
parameters  with the relation~\eqref{e:nutoa}  and to  the
$\partial_{\nu_i}Q_n$  using the relation~\eqref{e:QtoK}. The main
difference with the analysis in that paper lie in the fact that we
considering unordered integral expressions where the absence of
boundaries imply the vanishing of total derivative contributions. We
would like to stress that these identities are crucial for the
observed extra simplicity of unordered gravity amplitudes. It is
important that the crossing symmetry of gravity amplitude assure
that each ordered amplitude have the same tensorial structure
leading to the generic structure for the amplitude as given in
eq.~(\ref{e:cPsusy}) allowing us to use the reduction formulas
derived in that section.

These rules are the general cases of the identities used in
ref.~\cite{BjerrumBohr:2008vc} for the cancellation of triangle
contributions to the five graviton amplitude at one-loop in $\cN=8$
supergravity.

%%%%%%%%%%%%%%%%%%%%%%%%%%%%%%%%%%%%%%%%%%%%%%%%%%%%%%%%%%%%%%%%%%%%%%%%%%%%
\section{\label{sec:level3} Structure of the supergravity amplitude}
%%%%%%%%%%%%%%%%%%%%%%%%%%%%%%%%%%%%%%%%%%%%%%%%%%%%%%%%%%%%%%%%%%%%%%%%%%%%

The $n$-graviton amplitudes in  supergravity with $\cN$
supersymmetries at one-loop have the following
representation that we described in section~\ref{sec:level1b}
\begin{equation}\label{e:DefAmp}
\mathcal{M}^\cN_{n;1} =\Gamma(n-D/2)\times \int_0^1 d^{n-1}\nu
P(\varepsilon_{ij},k_i,\nu_i)\, Q_n^{D/2-n}\,.
\vspace{-.2cm}\end{equation}
Here $\varepsilon_{ij}=(h_i\bar h_j+h_j\bar h_i)/2$ are the
polarisations of the gravitons and the integral contains a
polynomial of order $2n-\cN$ in $\nu_i$ given by
$P(\varepsilon_{ij},k_i,\nu_i)\!=\! P(H_i\cdot K_{[n]},
Y_{ij}\,G_F(\nu_i-\nu_j), (h_i\cdot h_j)\, \delta(\nu_i-\nu_j))$,
where
\begin{equation}
\label{eq:K} K_{[n]}=\sum_{m=1}^n k_m\, \nu_m\,,
\end{equation}
is the total loop momentum, $H_i$ is $h_i$ or $\bar h_i$ and
$Y_{ij}\in\{(H_i\cdot H_j),(H_i\cdot  k_j),(k_i\cdot  k_j)\}$ for
$i\neq j$. The structure of $P(\varepsilon_{ij},k_i,\nu_i)$ can
easily be determined using the string  based
rules~\cite{StringBased,Strassler:1992zr,BjerrumBohr:2008vc} and is
given by bosonic contractions, plane wave factor contractions and
fermionic contractions as described in section~\ref{sec:level1b}. We
will discuss these in turn [we refer to the appendix~A
of~\cite{BjerrumBohr:2008vc} for conventions and derivation of these
field theory limits].

\medskip \noindent{$\bullet$\ \bf The bosonic contractions for $i\neq
j$}\vspace{-0.1cm}\smallskip
\begin{eqnarray}
\label{e:HigherD}
  \langle h_i\cdot \partial X(\nu_i) \, h_j\cdot
 \partial      X(\nu_j)\rangle       &=&      2(h_i\cdot      h_j)\,
 (\delta(\nu_i-\nu_j)-1)\,,\\
\nonumber \langle h_i\cdot \partial X(\nu_i)\,\bar h_j\cdot
\bar\partial X(\nu_j)\rangle&=&  2(h_i\cdot \bar h_j)\,
(\delta(\nu_i-\nu_j)-1)\,.
\end{eqnarray}
The    $\delta$-function  leads to   the   reducible contributions
which are present in the amplitude from five-point
order~\cite{BjerrumBohr:2008vc}.  These  give  rise  to  the
higher-point vertices to the field theory loop amplitude
contribution. The zero mode of the bosonic world-sheet coordinate
contributes to the constant  piece. This piece once plugged into the
first  contribution in eq.~(\ref{e:Feven})  leads to  a dimension
shifted integral after having integrated over the proper-time $T$ in
eq.~(\ref{e:AmpDef}). An amplitude with $\cN$ supersymmetries
contains the contributions $\cI_n^{[D+2k]}[\I_r,\J_s]$    for $0\leq
k\leq n-\cN/2$. [This corresponds to the contributions
$\mathcal{A}_n^{(2)\infty}$ in eq.~(2.7) and eq.~(5.4) of
reference~\cite{BjerrumBohr:2008vc}].
These contributions are proportional to $h_i\cdot \bar h_j$ for
$i\neq j$,   which   is   not   invariant  under   the (linearised)
gauge transformations  $h_i\to h_i+k_i\lambda$. Gauge  invariance is
recovered when momentum dependent contributions from integrating by
parts in the reduction formulas eq.~(\ref{e:DD}) are taken into
account.

\medskip\noindent{$\bullet$} {\bf The contraction with plane-wave
factors}%\vspace{-0.5cm}\smallskip
\begin{eqnarray}
\label{eq:hk2}
&&  \sum_{1\leq m\leq n\atop m\neq i } \, (h_i\cdot k_m)\,
\partial_{\nu_i} G_B(\nu_i-\nu_m)%\\
 = -2 h_i\cdot K_{[n]} - \sum_{m=1}^n (h_i\cdot
k_m)\,G_F(\nu_i-\nu_m)\,.%\vspace{-0.3cm}
\end{eqnarray}
Here we have used that a given external state has to be contracted
with all the plane-wave factors of the other external states.

\smallskip\medskip\noindent{$\bullet$} {\bf The fermionic contractions for
$i\neq j$}\vspace{-0.1cm}\smallskip
\begin{equation}
\label{eq:hhF} \langle v_i\cdot \psi(\nu_i) \, v_j\cdot
\psi(\nu_j)\rangle= (v_i\cdot v_j)\,    G_F(\nu_i-\nu_j)\,.
\end{equation}
Here $v_i$ stands for an external polarisation $h_i$ or momentum
$k_i$.\medskip

In $D$ dimensions we can  expand each of the polarisation tensors
$h_i$ and $\bar h_i$ in a basis of independent
momenta\vspace{-0.1cm}
\begin{equation}\label{e:cij}
h_i=\sum_{j=1}^{n-1} c_i{}^j \, k_j+ q^\perp\,.\vspace{-0.2cm}
\end{equation}
Here  $q^\perp$ is  a  vector  orthogonal to  the $(n-1)$ linearly
independent external momenta $k_j$. We use an identical definition
for the $\bar h_i$ polarisations.
 Using that $\partial_{\nu_i}Q_n = - 2k_i\cdot K_{[n]}-\sum_{m=1}^n
(k_i\cdot k_m)\, G_F(\nu_i-\nu_m)$ one gets that\vspace{-0.2cm}
\begin{equation}\label{e:KtoQ}
h_i\cdot K_{[n]}=\!\sum_{j=1}^{n-1}\! {c_i{}^j\over2}
\!\Big[-\partial_{\nu_j}Q_n\!+\!\! \sum_{m=1}^n\! (k_j\cdot
k_m)G_F(\nu_j-\nu_m)\Big]\,. \vspace{-0.3cm}\end{equation}
Hence the amplitude eq.~(\ref{e:DefAmp}) reduces to a sum of
integral contributions of eq.~(\ref{e:DefIn}) involving
$\partial_\nu Q_n$ and $G_F$ factors\vspace{-0.1cm}
\begin{equation}\label{e:dec}
\mathcal{M}_{n;1}^{\cN}=   \sum_{r+s+u=2n-\cN\atop 0\leq u\leq n}\sum_{l=0}^u\, t^l_{r,s}\,\cI^{[D+2(u-l)]}_{n-l}[\I_r,\J_s]\,.
\vspace{-0.2cm}\end{equation}
In this equation $t_{r,s}$ is the tensorial structure to be
discussed in more details elsewhere~\cite{BjbohrVanhove}, and
$\cI_n[\I_r,\J_s]$ are the loop integrals on which we can apply the
reduction formulas in eq.~(\ref{e:Red1}) and eq.~(\ref{e:Red2}).

%%%%%%%%%%%%%%%%%%%%%%%%%%%%%%%%%%%%%%%%%%%%%%%%%%%%%%%%%%%%%%%%%%%%%%%%%%%%
\section{\label{sec:level4}MAXIMAL SUPERGRAVITY}
%%%%%%%%%%%%%%%%%%%%%%%%%%%%%%%%%%%%%%%%%%%%%%%%%%%%%%%%%%%%%%%%%%%%%%%%%%%%

For the case of $\cN=8$ supergravity the highest contribution of
loop momentum is for $r=2(n-4)$ and $s=0$. There are  no powers of
$G_F$ in the integral. A direct application of $n$ steps of the
reduction rules eq.~(\ref{e:Red1}) gives% \vspace{-0.3cm}
\begin{equation}\begin{split}
\cI_{n}[(\partial Q_n)^{2(n-4)}]&\rightsquigarrow
\cI_{n-1}^\m[(\partial Q_n)^{2(n-5)}] + \cI_n^{[D+2]}[(\partial
Q_n)^{2(n-5)}]\cr & \rightsquigarrow \cdots \rightsquigarrow
\cI_4^\m[\emptyset]+\sum_{m=1}^{n-4}
\cI_{4+m}^{[D+2m]}[\emptyset]\,.
\end{split}\vspace{-0.4cm}\end{equation}
Hence we end up with scalar box integral functions plus dimension
shifted scalar integrals. We demonstrate that these cancel when one
includes the dimension shifted contribution from the contractions of
polarisations in eq.~(\ref{e:HigherD}).

For $s\neq 0$, and $r>s$ one first applies $s$ steps of the
reduction formula eq.~\REF{e:Red2} to get\vspace{-0.1cm}
\begin{equation}
\cI_{n}[(\partial Q_n)^{r},(G_F)^s]\rightsquigarrow
\cI_{n-s}^\m[(\partial Q_n)^{r-s}]+\cI_{n}^{[D+2]}[(\partial Q_n)^{r-s}]\,,
\vspace{-0.1cm}\end{equation}
then one applies $(r-s)/2$ steps of the reduction formula
eq.~\REF{e:Red1} to get to\vspace{-0.1cm}
\begin{equation}\begin{split}
\cI_{n}[(\partial Q_n)^{r},(G_F)^s]&\rightsquigarrow
\cI_{n-s}^\m[(\partial Q_n)^{r-s}]+\cI_{n}^{[D+2]}[(\partial Q_n)^{r-s}]\rightsquigarrow\cdots\cr &
\rightsquigarrow
\cI_{n-(r+s)/2}^\m[\emptyset]+\sum_{m=1}^{n-4}\cI_{4+m}^{[D+2m]}[\emptyset]\,.
\end{split}\vspace{-.3cm}\end{equation}
Since $r\!+\!s\!=\!2(n-4)$ in $\cN=8$ supergravity one ends with
scalar box integral functions plus dimension shifted scalar
integrals which again will cancel in the total amplitude. For $s>r$
there are only vanishing contributions from the reduction formulas
via eq.~(\ref{e:noHD}) and its generalisation.

\subsection{Cancellation of the dimension shifted integrals}\label{sec:dimshift}
 The dimension shifted scalar
integrals generated by the reduction formula have the momentum space representation
\begin{eqnarray}\label{e:DD}
&&  \cI^{[4+2m]}_{4+m}[\emptyset]= -i {(-1)^k (4\pi)^{2+m-\epsilon}\over
    \Gamma(2+\epsilon)}\, \times%\\
%\nonumber&\times&
\int_0^\infty {d^{4-\epsilon}\ell
    d^{2m}\ell_\perp\over                    (2\pi)^{4+2m-2\epsilon}}\,
  \prod_{i=1}^{4+m}{1\over (\ell-k_1-\cdots k_i)^2+\ell_\perp^2}\,.\vspace{-.3cm}
\end{eqnarray}
The $4+m$-point scalar loop  integral with four dimensional
kinematics in eq.~(\ref{e:DD}) is evaluated  in $4+2m$ dimensions,
does not carry any UV or IR divergences and is finite when
$\epsilon\to0$.

The reduction formulas derived in  the previous section do not contain
any  information  about  the   gauge  invariance  and  the  number  of
supersymmetries  of the theory.  In the  total physical  amplitude the
higher dimensional contributions generated by these reduction formulas
will   combine    with   the    ones   from   the    contractions   in
eq.~(\ref{e:HigherD})  so  that   they  appear  with  gauge  invariant
coefficients, as  we show  below. For $\cN=8$  supergravity amplitudes
these contributions sum up to  total derivatives and do not contribute
to the physical amplitude.

For two given external states labelled $i$ and $j$ and with $H_i$
for the polarisations $h_i$ or $\bar h_i$, the zero-mode part of
each contraction in eq.~(\ref{e:HigherD}) contributes to the
amplitude as follows
\begin{eqnarray}
\nonumber \langle H_i\cdot \partial_{\nu_i}X \, H_j\cdot\bar\partial_{\nu_j}
X\rangle&\to& -2(H_i\cdot H_j)\, Q_n\\
\label{e:hhh}&=&-2\, Q_n\, \sum_{r=1}^{n-1}\sum_{s=1}^{n-1}\,c_i{}^r  c_j{}^s\, (k_r\cdot
k_s)\\
\nonumber  &=& \sum_{r=1}^{n-1}\sum_{s=1}^{n-1}\,c_i{}^r   c_j{}^s\,
Q_n\, [\partial_{\nu_r}\partial_{\nu_s}Q_n- 2\delta(\nu_r-\nu_s)]\,.
\end{eqnarray}
In the second line we have used the expansion of the polarisations
in the basis of momenta defined in eq.~(\ref{e:cij}). The dimension
shifted integrals arises from the two derivatives acting on $Q_n$.
We will now show that this contribution will  cancel against a
corresponding contribution from the  contractions of  the
polarisations with  the plane-wave factors in eq.~(\ref{eq:hk2})
\begin{eqnarray}
\label{e:hkk}\!\sum_{1\leq p\leq n\atop p\neq i}\sum_{1\leq q\leq
n\atop q\neq j}\!\!\,\langle h_i\cdot \partial_{\nu_i}X\, k_p\cdot
X\rangle\,\langle \bar h_j\cdot\bar\partial X_{\nu_j}    \,k_q\cdot
X\rangle&\to&    \sum_{r=1}^{n-1}   \sum_{s=1}^{n-1} c_i{}^r c_j{}^s
\partial_{\nu_r}Q_n\partial_{\nu_s}Q_n\,.
\end{eqnarray}
Since the contributions eq.~(\ref{e:hhh}) corresponds to the
contraction  between  two  polarisation of  the  external  states
and eq.~(\ref{e:hkk})   corresponds  to   the   contraction  between
the polarisations and the plane-wave factors these contributions
arise in the amplitude with the following coefficients
\begin{eqnarray}
  {\rm eq.}\ (\ref{e:hhh})+{\rm eq.}~(\ref{e:hkk})&\to& \sum_{r=1}^{n-1}   \sum_{s=1}^{n-1}
c_i{}^r c_j{}^s \, [\partial_{\nu_r}Q_n\partial_{\nu_s}Q_n+
Q_n\,                              \partial_{\nu_r}\partial_{\nu_s}Q_n-
2Q_n\,\delta(\nu_r-\nu_s)]\\
\nn &=& \sum_{s=1}^{n-1} c_i{}^r c_j{}^s \,
[\partial_{\nu_r}(Q_n\partial_{\nu_s}Q_n)-
2Q_n\,\delta(\nu_r-\nu_s)]\,.
\end{eqnarray}
The second given by the delta-function collapses two external  legs
and generates a massive  scalar integral  with  one leg less. The
first term generated by  integration by parts produces a dimension
shifted integral of the  same structure as the one generated by the
reduction formula in eq.~(\ref{e:R1}).

 We will now show how this works in few specific
examples.

\medskip\noindent{$\bullet$}  At  five  point  order  the  $\cN=8$
amplitude in eq.~(\ref{e:dec}) involves  only $\cI_5[(\partial
Q_5)^r,(G_F)^s]$ with $r+s=2$. The contribution $\cI_5[(\partial
Q_5)^2]$ arises from the contractions of eight left-moving and
right-moving fermions and corresponds to  the contributions
$\mathcal{A}^{(2)\infty}_5$ and the contribution
$|\mathcal{A}^{(1)\infty}_5-t_{10}\cdot F^5|^2$ in eq.~(2.7) of
ref.~\cite{BjerrumBohr:2008vc}

\begin{equation}
  \delta\mathcal{M}^{\cN=8}_{5;1}=\sum_{i,j=1}^5\,                 t_8
  F^4_{\hat\imath}t_8F^4_{\hat\jmath}\,  \langle   (h_i\cdot\partial  X)\,  (\bar
  h_j\cdot \bar\partial X)\, \prod_{r=1}^5 e^{ik_r\cdot X}
  \rangle\,,
\end{equation}
where    $    t_8F^4_{\hat\imath}$    is    defined    in eq.~(2.1)
of ref.~\cite{BjerrumBohr:2008vc}. Focusing  on  the contractions
eq.~(\ref{e:hhh})  and eq.~(\ref{e:hkk}),  this expression leads to
the following contributions
\begin{eqnarray}
\delta \mathcal{M}^{\cN=8}_{5;1}\Big|_{(\ref{e:hhh})+(\ref{e:hkk})}&=&\sum_{i,j=1}^5\,                 t_8
  F^4_{\hat\imath}t_8F^4_{\hat\jmath}\,   c_i{}^r   c_j{}^s   \,
  \times\\
\nonumber &\times&\Gamma(5-D/2)\,\int_0^1 d^4\nu \,
Q_5^{D/2-5}\,[\partial_{\nu_r}Q_5\partial_{\nu_s}Q_5+{1\over     D/2-4}
Q_5 \partial_{\nu_r}\partial_{\nu_s}Q_5]\\
\nonumber
&=&\sum_{i,j=1}^5\,                 t_8
  F^4_{\hat\imath}t_8F^4_{\hat\jmath}\,   c_i{}^r   c_j{}^s   \,
\Gamma(3-D/2)\,\int_0^1 d^{4}\nu
\,\partial_{\nu_r}\partial_{\nu_s}Q_5^{D/2-3}\,,
\end{eqnarray}
which again leads to a vanishing total derivative by the
1-periodicity of $Q_5$ with respect to each of the $\nu_i$
variables.  This implies that in the five-point $\cN=8$ supergravity
amplitude  the highest contribution  in  loop momentum
$\cI_5[(\partial  Q_5)^2]$ has been completely  cancelled against
the  dimension shifted contribution $\cI_5[Q_5\partial^2Q_5]$.   The
other contribution $\cI_5[\partial Q_5,G_F]$ does not receive
contributions  from the  dimension  shifted integral (thanks to
eq.~(\ref{e:noHD}))  and is hence completely reducible to scalar box
integrals defined in $D=4-2\epsilon$ dimensions. $\cI_5[(G_F)^2]$ is
vanishing by the generalisation of eq.~(\ref{e:noHD}).

\medskip\noindent{$\bullet$} The six  point $\cN=8$ amplitude
in eq.~(\ref{e:dec}) involves  the  contributions  $\cI_6[(\partial
Q_6)^r,(G_F)^s]$  with $r+s=4$. The contribution $\cI_6[(\partial
Q_6)^4]$ arises from the contractions of eight  left-moving and
right-moving fermions [from the square of the last  term in
$\mathcal{A}_6^{(1)}$ in eq.~(5.1) and the last      contribution to
$\mathcal{A}_6^{(2)}$      in eq.~(5.4) in
ref.~\cite{BjerrumBohr:2008vc}] and is given by
\begin{eqnarray}
 \delta \mathcal{M}_{6;1}^{\cN=8}= \sum_{ij,pq} \, t_{ij,pq}\,
 \langle (h_i\cdot \partial X) (h_j\cdot \partial X)\, (\bar h_p\cdot\bar\partial X)\,
 (\bar h_q\cdot \bar\partial X)\, \prod_{r=1}^6 e^{ik_r\cdot X}
 \rangle\,,
\end{eqnarray}
where $t_{ij,pq}=t_8F^4_{\hat\imath,\hat\jmath}t_8F^4_{\hat p,\hat
q}$, and with $t_8F^4_{\hat\imath,\hat\jmath}$ defined as in
ref.~\cite{BjerrumBohr:2008vc} [It is a tensorial structure composed
from  four of the external  polarisations and four of the external
momenta]. The $H_i\cdot  H_j$ contractions and the contractions with
the plane-wave factor $\exp(i k\cdot X)$ lead to
\begin{eqnarray}\hspace{-0.2cm}
  \delta_1 \mathcal{M}_{6;1}^{\cN=8}&\!=\!&     \sum_{r,s,t,u}     \,    t^{rstu}\,
  \Gamma(6-D/2)\,\int_0^1         d^5\nu\,         Q_6^{D/2-6}\, \times
\Big[\partial_{\nu_r}Q_6\partial_{\nu_s}Q_6\,\partial_{\nu_t}Q_6\partial_{\nu_u}Q_6\\
\nonumber&+&
 {Q_6^2\over (D/2-4)(D/2-5)}\, \big(\partial_{\nu_r}\partial_{\nu_s}Q_6
\,\partial_{\nu_t}\partial_{\nu_u}Q_6+ \partial_{\nu_r}\partial_{\nu_u}Q_6
\,\partial_{\nu_t}\partial_{\nu_s}Q_6+
\partial_{\nu_r}\partial_{\nu_t}Q_6
\,\partial_{\nu_s}\partial_{\nu_u}Q_6
\big)\\
\nonumber&+&                     {Q_6\over                    D/2-5}\,
\big(\partial_{\nu_r}Q_6\partial_{\nu_s}Q_6\, \partial_{\nu_t}\partial_{\nu_u}Q_6+\partial_{\nu_t}Q_6\partial_{\nu_u}Q_6\, \partial_{\nu_r}\partial_{\nu_s}Q_6+\partial_{\nu_s}Q_6\partial_{\nu_u}Q_6\, \partial_{\nu_r}\partial_{\nu_t}Q_6\\
\nonumber&+&
\partial_{\nu_t}Q_6\partial_{\nu_s}Q_6\, \partial_{\nu_r}\partial_{\nu_u}Q_6+\partial_{\nu_r}Q_6\partial_{\nu_u}Q_6\, \partial_{\nu_t}\partial_{\nu_s}Q_6+
\partial_{\nu_r}Q_6\partial_{\nu_t}Q_6\, \partial_{\nu_s}\partial_{\nu_u}Q_6
\big)\Big]\,.
\end{eqnarray}
Here  we have  defined $t^{rstu}\equiv t_{ijpq} \, c_i{}^r c_j{}^t
c_p{}^s c_q{}^u$. Using that the third (and higher) derivative term
$\partial_{\nu_i}\partial_{\nu_j}\partial_{\nu_k}Q_n=0$ vanishes, we
can rewrite the previous expression as a  total derivative
\begin{eqnarray}
 \delta_1  \mathcal{M}_{6;1}^{\cN=8}&=&     \sum_{r,s,t,u}     \,   t^{rstu}\,
  \Gamma(4-D/2)\,\int_0^1         d^5\nu\,
\,
\partial_{\nu_r}\partial_{\nu_s}\big(Q_6^{D/2-4} \partial_{\nu_t}Q_6\partial_{\nu_u}Q_6\big)\\
\nonumber&+&   \sum_{r,s,t,u}     \,   t^{rstu}\,
  \Gamma(4-D/2)\,\int_0^1 d^5\nu\,\partial_{\nu_r}(Q_6^{D/2-4}\,
\,\partial_{\nu_s}Q_6\partial_{\nu_t} \partial_{\nu_u}Q_6)\,.
\end{eqnarray}
As before the boundary terms  vanish by the 1-periodicity of the
integrand.    Therefore   the   highest   power   of   loop momentum
$\cI_6[(\partial Q_6)^4]$ in the $\cN=8$ supergravity six-point
amplitude has completely  cancelled against the dimension shifted
contributions.

The  contribution  $\cI_6[(\partial  Q_6)^3,G_F]$  combines  with
the dimension shifted integrals according to
\begin{eqnarray}
\delta_2\mathcal{M}_{6;1}^{\cN=8}&=&\sum_{r,s,u} t^{rsu} \,\Gamma(6-D/2)\, \int_0^1 d^5\nu Q_6^{D/2-6}\, G_F(\nu_p-\nu_q)\,
[\partial_{\nu_r}Q_6 \partial_{\nu_s}Q_6\partial_{\nu_u}Q_6 \\
\nonumber &+&
{Q_6\over(D/2-5)}                                                    \,
(\partial_{\nu_r} \partial_{\nu_s}Q_6\partial_{\nu_u}Q_6
+\partial_{\nu_r} \partial_{\nu_u}Q_6\partial_{\nu_s}Q_6+\partial_{\nu_r}Q_6\partial_{\nu_s}\partial_{\nu_u}Q_6)
]\\
\nonumber &=&-\sum_{r,s,u} t^{rsu} \,\Gamma(3-D/2)\, \int_0^1 d^5\nu \, G_F(\nu_p-\nu_q)\,
\partial_{\nu_r}\partial_{\nu_s}\partial_{\nu_u}Q_6^{D/2-3}\\
\nonumber  &=&2\sum_{s,u} (t^{psu}-t^{qsu}) \,\Gamma(3-D/2)\, \int_0^1 d^5\nu \, \partial_{\nu_s}\partial_{\nu_u}(\delta(\nu_p-\nu_q)\delta(\nu_r-\nu_p)\,
Q_6^{D/2-3})\,,
\end{eqnarray}
where $\nu_p$ and $\nu_q$ denotes positions of external states. This
expression vanishes  as a total derivative.  The dimension shifted
contributions in the quadratic hexagon $\cI_6[(\partial
Q_6)^2,(G_F)^2]$ are treated in a similar fashion as follows
\begin{eqnarray}
\nonumber \delta_3\mathcal{M}_{6;1}^{\cN=8}&=&\sum_{r,s} t^{rs}
\,\Gamma(6-D/2)\, \int_0^1 d^5\nu Q_6^{D/2-6}\, \prod_{x\in\J_2} G_F(x)\,
[\partial_{\nu_r}Q_6 \partial_{\nu_s}Q_6+
{Q_6\over(D/2-5)}                                                    \,
\partial_{\nu_r} \partial_{\nu_s}Q_6
]\\
\nonumber &=&\sum_{r,s} t^{rs} \,\Gamma(4-D/2)\, \int_0^1 d^5\nu \,\prod_{x\in\J_2} G_F(x)\,
\partial_{\nu_r}\partial_{\nu_s}Q_6^{D/2-4}\\
\nonumber    &=&-2\sum_{s}     \,\Gamma(3-D/2)\,    \int_0^1    d^5\nu
\, \partial_{\nu_s}Q_6^{D/2-4}\\
&\times&\,[ (t^{ps}-t^{qs})\delta(\nu_p-\nu_q)
G_F(\nu_u-\nu_v)+ (t^{us}-t^{vs}) \delta(\nu_u-\nu_v)
G_F(\nu_p-\nu_q)]\,,
\end{eqnarray}
and is given by  a sum of linear one mass  pentagons, which are
completely reducible to scalar boxes.  The last two contributions
are the linear hexagon $\cI_6[\partial Q_6,(G_F)^3]$ and  the scalar
hexagon $\cI_6[(G_F)^4]$ which are vanishing.

We have thus shown  that the dimension shifted contributions from
the contractions in eq.~(\ref{e:HigherD}) cancel  against the ones
from the reduction formulas.    We  would like  to stress that  this
mechanism does not require any supersymmetry.

\smallskip \begin{center}\emph{ This  implies that the only basis  functions for the
  one-loop   $n$-graviton   amplitude\\   in   $\cN=8$  supergravity   are
  four-dimensional scalar box integral functions. \\ This proves the
  ``no-triangle'' hypothesis. }\end{center}

\newpage
%%%%%%%%%%%%%%%%%%%%%%%%%%%%%%%%%%%%%%%%%%%%%%%%%%%%%%%%%%%%%%%%%%%%%%%%%%%%
\section{\label{sec:level5} GRAVITY with less or no supersymmetries}
%%%%%%%%%%%%%%%%%%%%%%%%%%%%%%%%%%%%%%%%%%%%%%%%%%%%%%%%%%%%%%%%%%%%%%%%%%%%

For less supersymmetry or in the case of pure gravity the
application of the reduction formula eq.~(\ref{e:Red1}) and
eq.~(\ref{e:Red2}) leads to\vspace{-0.25cm}
\begin{equation}
\cI_n[(\partial Q_n)^r,(G_F)^s]\to \cI^\m_{\cN/2}[\emptyset]\,.
\vspace{-0.1cm}\label{genericN}\end{equation}
The endpoint of the chain of reductions is given by a scalar
$\cN/2$-point integral function. The dimension shifted integrals
cancel in the physical amplitude following the previous arguments.

\medskip \noindent We can conclude the following from eq.~(\ref{genericN}):

\medskip \noindent{$\bullet$} Theories with $\cN\geq 3$ contain
integral functions down to scalar bubbles and are hence one-loop cut
constructible. This confirms the analysis of ref.~\cite{Bern:2007xj}
from considerations of on-shell unitarity cuts of the amplitude.

\medskip \noindent{$\bullet$} For $\cN=0$  gravity: one-loop
amplitudes are reducible down to rational parts  as are QCD
amplitudes. Since in four dimensions only  the two-point (bubble)
integral has  UV logarithmic divergences, we conclude that one-loop
gravity amplitudes are  at most logarithmically  diverging. For pure
gravity this divergence cancels on-shell~\cite{'t Hooft:1974bx} but
is present when coupled to matter~\cite{Deser:1974cz}.

%%%%%%%%%%%%%%%%%%%%%%%%%%%%%%%%%%%%%%%%%%%%%%%%%%%%%%%%%%%%%%%%%%%%%%%%%%%
\section{\label{sec:level6}DISCUSSION AND CONCLUSIONS}
%%%%%%%%%%%%%%%%%%%%%%%%%%%%%%%%%%%%%%%%%%%%%%%%%%%%%%%%%%%%%%%%%%%%%%%%%%%%

In this paper we have  explored the integral expansion of the
one-loop $n$-point  graviton amplitude  in pure  gravity and  in
supersymmetric extensions.  It has  been shown  that these unordered
amplitudes are constrained by  new integral  reduction formulas for
colourless gauge theories in  four dimensions. Decomposing the
polarisation tensors in the amplitudes  in a basis of  independent
momenta enables  the use of these reduction formulas in a form that
does not  require the need to invert  any Gram determinant  in the
kinematic  variables. The  Gram determinant generally vanishes  for
special  kinematic configurations associated with  particle
productions   at  thresholds   or  planar dependence between
external momenta. This is particularly suitable for a numerical
analysis   of   the  $\epsilon$   expansion   of   the
amplitude~\cite{BjbohrVanhove}. For  maximal $\cN=8$ supergravity
this leads  to  ``no-triangle'' properties  of  the $n$-point
supergravity amplitudes. This shows that the $n$-graviton amplitude
at one-loop is completely   specified  by scalar   box integral
functions. The  proof  of  the  'no triangle hypothesis'  of
one-loop amplitudes in $\cN=8$  supergravity was mainly discussed in
the case of external graviton states but is generalisable to all
matter states in the supergravity multiplet. This is because the
constraint eq.~\eqref{e:rsConstraint} arising from the cancellation
of $\cN$  powers of  loop momenta in the loop amplitude do not
depend on the nature of the external states due to supersymmetry.
Because the integrals $\cI_n[\I_r,\J_s]$ which arise from the
correlations of the world-line fields in eq.~\eqref{e:HigherD1} and
eq.~\eqref{eq:hhF1} have the same form whatever  the external
massless states,  any one-loop $n$-point amplitude between states of
the massless supergravity multiplet will lead thus to amplitudes of
the form eq.~\eqref{e:cPsusy}. For these amplitudes the unordered
reduction formula can be directly applied. In $\cN=4$ supergravity
it means that the $n$-graviton amplitude contains only integral
functions  up to scalar bubbles and thus is   constructible  from
its cuts in $D=4-2\epsilon$.  For pure gravity  our  result  yields
an amplitude consisting of  scalar box, triangle and bubble
integrals as  well as rational pieces.

The lack of colour in massless QED, means that one can as well apply
the reduction formulas eq.~(\ref{e:Red1}) and eq.~(\ref{e:Red2}) to
the light-light $n$-photon scattering at one-loop. For instance, a
six-photon one-loop amplitude has at most six powers of loop
momenta, $\cI_6[\I_r,\J_s]$ with $r+s=6$. Here the reduction
formulas imply that the four dimensional one-loop amplitude can be
expanded in terms of scalar box and triangle integral functions in
$D=4-2\epsilon$ dimensions. This is in agreement with the results
of~\cite{Binoth:2007ca,Bernicot:2007hs}.

Because of the colour factors in Yang-Mills theory, $n$-gluon
amplitudes in $\cN=4$ super-Yang-Mills have to be reduced using the
usual colour-ordered reduction
formulas~\cite{Bern:1993kr,Bern:2005hh}. Here only one power of loop
momentum is cancelled at each step of reduction and the $n$-point
amplitude is completely determined by scalar box integral functions.
It is interesting to note the following: in colourless theories the
sum over all orderings exactly produces the extra cancellations of
loop momenta required to arrive at the same structure of the
$n$-point one-loop $\cN=8$ supergravity amplitude as of the
$n$-point one-loop $\cN=4$ super-Yang-Mills amplitude. This directly
explains the similarity of the UV and IR structure in one-loop
$n$-point amplitudes in $\cN=8$ supergravity and $\cN=4$
super-Yang-Mills.

The absence of triangles in one-loop $\cN=8$ supergravity amplitudes
restricts the form of the multi-loop amplitudes~\cite{Bern:2006kd},
and is a necessary (but not sufficient) requirement for the absence
of the three-loop divergence in four dimensions~\cite{Bern:2007hh}
and the possible perturbative finiteness of $\cN=8$ supergravity in
four dimensions~\cite{BjerrumBohr:2006yw,
GreenGT,Bern:2006kd,GreenYU,Bern:2007xj}. The results of this paper
adds to the empirical knowledge of perturbative ${\cal N}=8$
supergravity and $\cN=4$ super-Yang-Mills. It would be interesting
to investigate if similar unordered integral reductions are possible
to employ in the analysis of higher loop integrals. If so this could
add another important clue in the of understanding of why
higher-loop amplitudes in ${\cal N}=8$ supergravity and $\cN=4$
super-Yang-Mills seemingly have the same
UV-behaviour~\cite{GreenGT,Bern:2006kd} in four dimensions.

\section*{Acknowledgments}
We would like to thank Ignatios Antoniadis, Zvi Bern, Lance Dixon,
Gia Dvali, Harald Ita, and  Pierpaolo Mastrolia for useful
discussions. We would  like  to particularly  thank  Lance  Dixon
for attracting our attention regarding  the QED case and stressing
its similitude to the gravitational case.  PV would like to thank
the theory division of CERN for its hospitality when this paper has
been written.   The research  of (NEJBB)  was supported by grant
DE-FG0290ER40542 of the US Department of Energy. The research of
(PV) was supported in part the RTN contracts MRTN-CT-2004-005104 and
by the ANR grant BLAN06-3-137168.


\begin{thebibliography}{99}



\bibitem{Bern:1993kr}
Z.~Bern, L.~J.~Dixon and D.~A.~Kosower,
  %``Dimensionally Regulated One Loop Integrals,''
  Phys.\ Lett.\  B {\bf 302} (1993) 299
  [Erratum-ibid.\  B {\bf 318} (1993) 649]
  [arXiv:hep-ph/9212308];
  %%CITATION = PHLTA,B302,299;%%
%  Z.~Bern, L.~J.~Dixon and D.~A.~Kosower,
 % ``Dimensionally regulated pentagon integrals,''
  Nucl.\ Phys.\  B {\bf 412}, 751 (1994)
  [arXiv:hep-ph/9306240].
  %%CITATION = NUPHA,B412,751;%%


\bibitem{intred}
  L.~M.~Brown and R.~P.~Feynman,
 % ``Radiative corrections to Compton scattering,''
  Phys.\ Rev.\  {\bf 85} (1952) 231.
  %%CITATION = PHRVA,85,231;%%
  G.~Passarino and M.~J.~G.~Veltman,
  % ``One Loop Corrections For $e^+e^-$ Annihilation Into $\mu^+\mu^-$
  % In The Weinberg Model,''
  Nucl.\ Phys.\  B {\bf 160} (1979) 151.
  %%CITATION = NUPHA,B160,151;%%

\bibitem{Campbell:1996zw}
  J.~M.~Campbell, E.~W.~N.~Glover and D.~J.~Miller,
  %``One-loop tensor integrals in dimensional regularisation,''
  Nucl.\ Phys.\  B {\bf 498} (1997) 397
  [arXiv:hep-ph/9612413].
  %%CITATION = NUPHA,B498,397;%%

\bibitem{Denner:2005nn}
  A.~Denner and S.~Dittmaier,
  %``Reduction schemes for one-loop tensor integrals,''
  Nucl.\ Phys.\  B {\bf 734} (2006) 62
  [arXiv:hep-ph/0509141].
  %%CITATION = NUPHA,B734,62;%%


\bibitem{Dunbar}
D.~C.~Dunbar and P.~S.~Norridge,
%``Infinities within graviton scattering amplitudes,''
Class.\ Quant.\ Grav.\  {\bf 14} (1997) 351 [hep-th/9512084].
%%CITATION = HEP-TH 9512084;%%
%
D.~C.~Dunbar and N.~W.~P.~Turner,
%``Gravity and form scattering and renormalisation of gravity in six and eight
%dimensions,''
Class.\ Quant.\ Grav.\  {\bf 20}, 2293 (2003) [hep-th/0212160].
%%CITATION = HEP-TH 0212160;%%



\bibitem{Bern:1998sv}
  Z.~Bern, L.~J.~Dixon, M.~Perelstein and J.~S.~Rozowsky,
 % ``Multi-leg one-loop gravity amplitudes from gauge theory,''
  Nucl.\ Phys.\  B {\bf 546}, 423 (1999)
  [arXiv:hep-th/9811140].
  %%CITATION = NUPHA,B546,423;%%

%\cite{Bern:2005bb}
\bibitem{Bern:2005bb}
  Z.~Bern, N.~E.~J.~Bjerrum-Bohr and D.~C.~Dunbar,
  %``Inherited twistor-space structure of gravity loop amplitudes,''
  JHEP {\bf 0505}, 056 (2005)
  [arXiv:hep-th/0501137].
  %%CITATION = JHEPA,0505,056;%%

\bibitem{BjerrumBohr:2005xx}
  N.~E.~J.~Bjerrum-Bohr, D.~C.~Dunbar and H.~Ita,
%  ``Six-point one-loop N = 8 supergravity NMHV amplitudes and their IR behaviour,''
  Phys.\ Lett.\  B {\bf 621} (2005) 183
  [arXiv:hep-th/0503102].
  %%CITATION = PHLTA,B621,183;%%

\bibitem{BjerrumBohr:2006yw}
  N.~E.~J.~Bjerrum-Bohr, D.~C.~Dunbar, H.~Ita, W.~B.~Perkins and K.~Risager,
%  ``The no-triangle hypothesis for N = 8 supergravity,''
  JHEP {\bf 0612} (2006) 072
  [arXiv:hep-th/0610043].
  %%CITATION = JHEPA,0612,072;%%

\bibitem{Bern:2007xj}
  Z.~Bern, J.~J.~Carrasco, D.~Forde, H.~Ita and H.~Johansson,
 % ``Unexpected Cancellations in Gravity Theories,''
  arXiv:0707.1035 [hep-th].
  %%CITATION = ARXIV:0707.1035;%%



\bibitem{StringBased}
Z.~Bern and D.~A.~Kosower,
%``Efficient Calculation Of One Loop QCD Amplitudes,''
Phys.\ Rev.\ Lett.\  {\bf 66}, 1669 (1991);
%%CITATION = PRLTA,66,1669;%%
%Z.~Bern and D.~A.~Kosower,
%``The Computation of loop amplitudes in gauge theories,''
Nucl.\ Phys.\ B {\bf 379}, 451 (1992);
%%CITATION = NUPHA,B379,451;%%
Z.~Bern,
%``A Compact representation of the one loop N gluon amplitude,''
Phys.\ Lett.\ B {\bf 296}, 85 (1992).
%%CITATION = PHLTA,B296,85;%%
Z.~Bern, D.~C.~Dunbar and T.~Shimada,
%``String based methods in perturbative gravity,''
Phys.\ Lett.\ B {\bf 312}, 277 (1993) [hep-th/9307001].
%%CITATION = HEP-TH 9307001;%%
D.~C.~Dunbar and P.~S.~Norridge,
%``Calculation of graviton scattering amplitudes using string based methods,''
Nucl.\ Phys.\ B {\bf 433}, 181 (1995) [hep-th/9408014].
%%CITATION = HEP-TH 9408014;%%

\bibitem{Strassler:1992zr}
  M.~J.~Strassler,
  ``Field theory without Feynman diagrams: One loop effective actions,''
  Nucl.\ Phys.\  B {\bf 385} (1992) 145
  [arXiv:hep-ph/9205205].
  %%CITATION = NUPHA,B385,145;%%



\bibitem{BjerrumBohr:2008vc}
  N.~E.~J.~Bjerrum-Bohr and P.~Vanhove,
  %``Explicit Cancellation of Triangles in One-loop Gravity Amplitudes,''
  JHEP {\bf 0804} (2008) 065
  [arXiv:0802.0868 [hep-th]].
  %%CITATION = JHEPA,0804,065;%%

\bibitem{BjbohrVanhove} N.~E.~J.~Bjerrum-Bohr and P.~Vanhove, Work in
progress.


\bibitem{'t Hooft:1974bx}
  G.~'t Hooft and M.~J.~G.~Veltman,
  %``One loop divergencies in the theory of gravitation,''
  Annales Poincare Phys.\ Theor.\  A {\bf 20} (1974) 69.
  %%CITATION = AHPAA,A20,69;%%

\bibitem{Deser:1974cz}
  S.~Deser and P.~van Nieuwenhuizen,
  %``One Loop Divergences Of Quantized Einstein-Maxwell Fields,''
  Phys.\ Rev.\  D {\bf 10} (1974) 401.
  %%CITATION = PHRVA,D10,401;%%



\bibitem{Britto:2004nc}
  R.~Britto, F.~Cachazo and B.~Feng,
%  ``Generalized unitarity and one-loop amplitudes in N = 4  super-Yang-Mills,''
  Nucl.\ Phys.\  B {\bf 725} (2005) 275
  [arXiv:hep-th/0412103].
  %%CITATION = NUPHA,B725,275;%%

\bibitem{Binoth:2007ca}
  T.~Binoth, G.~Heinrich, T.~Gehrmann and P.~Mastrolia,
  %``Six-Photon Amplitudes,''
  Phys.\ Lett.\  B {\bf 649} (2007) 422
  [arXiv:hep-ph/0703311].
  %%CITATION = PHLTA,B649,


\bibitem{Bernicot:2007hs}
  C.~Bernicot and J.~P.~Guillet,
  %``Six-Photon Amplitudes in Scalar QED,''
  JHEP {\bf 0801} (2008) 059
  [arXiv:0711.4713 [hep-ph]].
  %%CITATION = JHEPA,0801,059;%%


\bibitem{Bern:2005hh}
Z.~Bern, V.~Del Duca and C.~R.~Schmidt,
  %``The infrared behavior of one-loop gluon amplitudes at
  %next-to-next-to-leading order,''
  Phys.\ Lett.\  B {\bf 445} (1998) 168
  [arXiv:hep-ph/9810409];
  %%CITATION = PHLTA,B445,168;%%
  Z.~Bern, N.~E.~J.~Bjerrum-Bohr, D.~C.~Dunbar and H.~Ita,
  %``Recursive calculation of one-loop QCD integral coefficients,''
  JHEP {\bf 0511} (2005) 027
  [arXiv:hep-ph/0507019].
  %%CITATION = JHEPA,0511,027;%%




\bibitem{Bern:2006kd}
  Z.~Bern, L.~J.~Dixon and R.~Roiban,
%  ``Is N = 8 supergravity ultraviolet finite?,''
  Phys.\ Lett.\  B {\bf 644}, 265 (2007)
  [arXiv:hep-th/0611086].
  %%CITATION = PHLTA,B644,265;%%

\bibitem{Bern:2007hh}
  Z.~Bern, J.~J.~Carrasco, L.~J.~Dixon, H.~Johansson, D.~A.~Kosower and R.~Roiban,
  %``Three-Loop Superfiniteness of N=8 Supergravity,''
  Phys.\ Rev.\ Lett.\  {\bf 98}, 161303 (2007)
  [arXiv:hep-th/0702112].
  %%CITATION = PRLTA,98,161303;%%



\bibitem{GreenGT}
M.~B.~Green, J.~G.~Russo and P.~Vanhove,
%``Non-renormalisation conditions in type II string theory and maximal
%supergravity,''
JHEP {\bf 0702}, 099 (2007)
[arXiv:hep-th/0610299].
%%CITATION = JHEPA,0702,099;%%

\bibitem{GreenYU}
M.~B.~Green, J.~G.~Russo and P.~Vanhove,
%``Ultraviolet properties of maximal supergravity,''
Phys.\ Rev.\ Lett.\  {\bf 98}, 131602 (2007)
[arXiv:hep-th/0611273].
%%CITATION = PRLTA,98,131602;%%






\end{thebibliography}
\end{document}